\documentclass[iop]{emulateapj}
\usepackage{subfigure}
\usepackage{amsmath}
\usepackage{color}

\newcommand{\cs}{c_{\rm s}}

\newcommand{\ob}{{\bf \Omega} \cdot {\bf B}}
\newcommand{\ratio}{\dot{M}_p/\dot{M}}

\begin{document}

\title{The Influence of Magnetic Field Geometry \\ on the Formation of Close-in Exoplanets}

\author{Jacob B. Simon$^{1,2}$}

\email{jbsimon.astro@gmail.com}

\begin{abstract}
Approximately half of Sun-like stars harbor exoplanets packed within a radius of $\sim$0.3 AU, but the formation of these planets and why they form in only half of known systems are still not well understood. 
We employ a one-dimensional steady state model to gain physical insight into the origin of these close-in exoplanets.  We use Shakura \& Sunyaev $\alpha$ values extracted from recent 
numerical simulations of protoplanetary disk accretion processes in which the magnitude of $\alpha$, and thus the steady-state gas surface density, depends on the orientation of large scale magnetic fields with respect to the disk's rotation axis.  Solving for the metallicity as a function of radius, we find that for fields anti-aligned with the rotation axis, the inner regions of our model disk often falls within a region of parameter space not suitable for planetesimal formation, whereas in the aligned case, the inner disk regions are likely to produce planetesimals through some combination of streaming instability and gravitational collapse, though the degree to which this is
true depends on the assumed parameters of our model.  More robustly, the aligned field case always produces higher concentrations of solids at small radii compared to the anti-aligned case. In the in situ formation model, this bimodal distribution of solid enhancement leads directly to the observed dichotomy in exoplanet orbital distances.
\end{abstract} 


\altaffiltext{1}{Department of Space Studies, Southwest Research Institute, Boulder, CO 80302}
\altaffiltext{2}{Sagan Fellow}

\section{Introduction}
\label{introduction}

With the recent discovery of thousands of exoplanets, we now know that stars' harboring of planets
is not unique to our Solar System.  Not only are exoplanets ubiquitous throughout our galaxy, but equally 
commonplace is the diversity of these systems. Indeed, discoveries such as 
hot Jupiters and super-Earths have made our own Solar System quite definitively not representative.  

A prime example of systems that differ from the Solar System are those in which
a large number of planets are packed into orbits smaller than $\sim$0.3 AU, comprising
about 50\% of all Sun-like systems \citep{winn15}. 
What causes these systems to be so dynamically packed?  One possibility is migration:
planets are formed at larger orbital distances and then
driven inward through angular momentum exchange with disk gas  \cite[e.g.,][]{terquem07,kley12}.
Another possibility is in situ formation: planets form from gas and solids already present 
at small distances from the central star.  For this mechanism to work, very high concentrations of solids must be present
at small radii \citep{hansen12,chiang13}.

A question perhaps even more intriguing than the origin of these close-in exoplanets is
how only $\sim 50$\% of all Sun-like systems have ended up in this configuration.  If these planets
do form in situ, then there should be a bimodal distribution of protoplanetary disk properties, in 
which $\sim 50$\% of disks have conditions favorable for the formation of planets at small radii,
and the remaining 50\% lack these conditions.

In this letter, we show that the bimodal distribution of observed exoplanet orbital distances can be traced back to the orientation of large scale magnetic fields relative to the rotation axis of the protoplanetary disk from which
these planets spawned.  The orientation of this magnetic field fundamentally affects the accretion stresses in the disk and thus the gas surface density, which in turn affects how small solids drift through the disk and their concentration relative to the gas.  As we show in the following sections, our calculations suggest that planetesimal formation is strongly favored for $r \lesssim 0.3$AU in 
systems with magnetic fields oriented parallel to the disk's rotation axis.  Assuming a uniform distribution
of magnetic field orientations and in situ formation, this equates to 50\% of systems producing exoplanets at small radii. 

The outline of this letter is as follows.  We describe our steady-state semi-analytic model in Section~\ref{model}, and then
present our main results in Section~\ref{results}.  We wrap up with a brief discussion in Section~\ref{summary}.

\section{Model}
\label{model}

\subsection{Disk Structure}

Our model consists of a 1D steady state disk comprised of both gas and solids.
The steady-state gas surface density is set by the turbulent viscosity, 

\begin{equation}
\label{sigma_gas}
\Sigma_g = \frac{\dot{M}}{3\pi\nu},
\end{equation}

\noindent
where $\dot{M}$ is the steady-state mass accretion rate onto the central star, which we assume to be $10^{-8} M_{\sun}/$yr, 
and the turbulent viscosity is defined via the standard \cite{shakura73} $\alpha$ parameter, $\nu = \alpha \cs^2/\Omega$.
$\cs$ is the locally isothermal sound speed of the gas (defined via the temperature below), and $\Omega$ is the Keplerian angular velocity.

We take the temperature to be that of the minimum mass solar nebula model \cite[MMSN;][]{hayashi81},

\begin{equation}
T(r) = 280 \left(\frac{r}{{\rm AU}}\right)^{-1/2} {\rm K}.
\end{equation} 

We treat the solid particles, with radius $a_p$ and mass density $\rho_p$, as a fluid with surface density $\Sigma_p(r,t)$.  Without any particle sinks or sources, the evolution of 
this surface density is governed by the continuity equation,

\begin{equation}
\label{dust_continuity}
\frac{\partial \Sigma_p}{\partial t} + \frac{1}{r}\frac{\partial}{\partial r}\left(r \left[F_{\rm diff} + F_{\rm adv}\right] \right) = 0,
\end{equation}

\noindent
where $F_{\rm diff}$ is the radial flux of solids due to turbulent diffusion, and $F_{\rm adv}$ is the radial flux of solids due to aerodynamic coupling between the solids and gas. 
The metallicity is $Z~\equiv~\Sigma_p/\Sigma_g$, and in the limit $Z \ll 1$, the diffusive flux is written as \citep{clarke88},

\begin{equation}
\label{diff_flux}
F_{\rm diff} = -D_p \Sigma_g \frac{\partial Z}{\partial r},
\end{equation}

\noindent where $D_p$ is the particle diffusivity.  

We can write a form for the particle diffusivity in terms of the dimensionless stopping time of the solids, $\tau = t_{\rm stop} \Omega$ (where $t_{\rm stop}$ is the {\it dimensional} stopping time) and the gas diffusivity $D$ (the diffusivity that would apply to a trace gas species within the disk) as $D_p = D/\left(1+\tau^2\right)$ \citep{youdin07b}.  The gas diffusivity scales with the turbulent viscosity $\nu$, with a constant of proportionality $\xi$ that is sensitive to the structure of whatever turbulent process is giving rise to angular momentum transport (discussed more in Section~\ref{alpha}); $D = \xi\nu$.

The advective radial flux $F_{\rm adv}$ arises because of aerodynamic coupling between solid particles and the gas.  For $\tau \ll 1$, the strongly coupled solids move at the same speed as the local gas, given by the accretion rate $\dot{M}$ as $v_r = -\dot{M}/\left(2\pi r\Sigma_g\right)$. For general $\tau$, there is also a component of the radial drift relative to the gas due to a mismatch between the azimuthal velocities of the particles and gas \citep{weidenschilling77b}. Combining these effects, we express the advective radial flux as \citep{takeuchi02},

\begin{equation}
\label{vp}
F_{\rm adv} = \Sigma_p v_p = \Sigma_p \frac{\tau^{-1}v_r - \eta v_{\rm K}}{\tau + \tau^{-1}},
\end{equation}

\noindent
where $v_{\rm K}$ is the Keplerian velocity and

\begin{equation}
\label{eta}
\eta = -\frac{1}{2}\left(\frac{h}{r}\right)^2\left[\frac{{\rm d ln}\Sigma_g}{{\rm d ln} r} + (q-3)\right],
\end{equation}

\noindent
where $q$ is defined via the local radial slope of the vertical gas scale height, $h = \cs/\Omega$,

\begin{equation}
\label{h_over_r}
\frac{h}{r} \propto r^{q-1}.
\end{equation}

\noindent
For the assumed temperature profile, $q = 1.25$.

We define the stopping time of the solids in two separate ways, depending on the location in the disk.  Inside of the snow line at 2.7 AU, we follow \cite{birnstiel09,birnstiel12}
and set the maximum $\tau$ by the fragmentation velocity, $v_{\rm frag}$,

\begin{equation}
\label{tau_inside}
\tau = \frac{1}{3 \alpha_{\rm mid}}\left(\frac{v_{\rm frag}}{\cs}\right)^2,
\end{equation}

\noindent
where $\alpha_{\rm mid}$ is the value of $\alpha$ at the disk mid-plane, where (through settling) most of the solids are concentrated. For silicates $v_{\rm frag} \approx~1$~m/s \citep{blum08}, but this velocity depends
non-trivially on particle size \citep{blum08}. Here, we assume $v_{\rm frag} = 1$~m/s for the fiducial case and vary it in what follows.

Outside of the snow line, we assume that particles reach mm sizes.  Our reason for this assumption is two fold.  First, coagulation simulations have indicated that beyond the snow line, solids can grow to larger sizes as the ice present
in these solids allows for enhanced sticking and is less susceptible to fragmentation \citep{birnstiel12}.  Second, sub millimeter observations show that mm size grains are prevalent at large radii in many protoplanetary disk systems \cite[e.g.,][]{andrews12}. Thus, the snow line serves as a natural transition from smaller particles (their exact size depending on $\alpha_{\rm mid}$) to larger particles that are seen in observations, albeit at large radii.
Thus, $\tau$ outside of the snow line is

\begin{equation}
\label{tau}
\tau = \frac{\pi}{2} \frac{\rho_p}{\Sigma_g} a_p,
\end{equation}

\noindent
where $\rho_p = 2 {\rm g/cm}^3$ and $a_p = 1$mm. 

Prior to disk dispersal, the gas surface density in the inner disk evolves on a time scale of $\sim$ Myr, while the local radial drift time, $r/|v_p|$, of mm to cm-size particles is orders of magnitude shorter. We can thus assume a fixed gas profile, and compute the steady-state distribution of particles drifting radially through the gas from a large reservoir situated further out. Equation~(\ref{dust_continuity}) then becomes,

\begin{equation}
\label{concentration_differential}
r\Sigma_g D_p\frac{dZ}{dr} - r\Sigma_g v_p Z = k.
\end{equation}

\noindent
The constant $k$ is the radial mass flux of solids, $k = \dot{M}_p/(2\pi)$, which 
we parameterize via the ratio of solid to gas accretion rates, $\ratio$.  This ratio is a free parameter in our model, but we can estimate values based on taking equation~(\ref{concentration_differential}) in the limit
of large radial distances, assuming diffusion is weak to advection at these distances (which is borne out by our calculations), and assuming $\tau \ll 1$.   At large $r$, radial drift dominates over accretion with the gas (again supported by our calculations).  One can then rewrite Equation~(\ref{concentration_differential}) to calculate $\ratio$ as a function of $Z$, $\tau$, $\alpha$, ${\rm d ln}\Sigma_g/{\rm d ln}r$, and $q$.  The scenario we envision is a reservoir of gas at large radii with ISM values for the metallicity ($Z = 0.01$).  We take $\tau = 0.1$ and $\alpha = 10^{-3}$--$10^{-2}$  as reasonable estimates\footnote{$\alpha$ may be
higher at larger radial distances because the gas surface density drops below the ionization depth of FUV photons \cite[see e.g.,][]{perez-becker11b}.}  at large distances, and constant $\alpha$ results in  ${\rm d ln}\Sigma_g/{\rm d ln}r$~=~-1 for our disk. For these values we find $\ratio \approx 0.1$--$1$; our fiducial value is $\ratio = 0.3$ and we explore $\ratio = 0.1$ and $\ratio = 1$ in what follows.

We numerically integrate Equation~(\ref{concentration_differential}) to determine the metallicity as a function of radius, assuming that the concentration of solids goes to zero at the inner edge of our disk, which we take to be the dust destruction radius,  $r_{\rm in} = 0.03{\rm AU}$ (for the temperature structure assumed here); $Z(r_{\rm in}) = 0$.   Finally, at the snow line, we reduce $Z$ by a factor of 2, following the arguments in \cite{lodders03}.

\subsection{Bimodal $\alpha$ from the Hall Effect}
\label{alpha}

A key component to our model is the dependence of $\alpha$ on the relative orientation of the large scale, vertical (i.e., perpendicular to the disk plane) magnetic field to the disk's rotation axis.\footnote{We assume that there is some non-negligible vertical magnetic field present because as shown by \cite{bai13b}, \cite{simon13a}, \cite{simon13b}, observed accretion rates are attainable only with the inclusion of a vertical field.}  Numerical
simulations that include all three non-ideal magnetohydrodynamical (MHD) effects present in protoplanetary disks have shown that for $r \lesssim 30$--60 AU, the Hall effect leads to two different states of the disk
within these regions, depending on the value of $\ob$ \citep{bai15,simon15b}.  For aligned field and angular momentum vector (i.e., $\ob > 0$) and for the field strengths explored here (see below), $\alpha \sim 0.01$ within the inner disk, whereas in the anti aligned case (i.e., $\ob < 0$), $\alpha \sim 10^{-4}$--$10^{-3}$ \citep{bai15,simon15b}.

We use the quantified $\alpha$ values from these numerical simulations in order to construct $\alpha(r)$ for our model.  In particular, we fit a power-law to the $\alpha$ values in the work of \cite{simon15b}
assuming a uniform in radius ratio of mid-plane gas to vertical magnetic pressure, $\beta_z = 10^5$. We choose this particular value because $\beta_z = 10^5$ provides the largest sample of $\alpha$ values from recent simulations,
and for reasonable gas surface density profiles (such as in the MMSN), the resulting $\alpha$ values correspond to $\dot{M} = 10^{-8} M_{\sun}/$yr (though, here, we do the opposite and assume $\dot{M} = 10^{-8} M_{\sun}/$yr and then calculate $\Sigma_g$).

Our power law fits to the $\alpha$ values are given by 

\begin{equation}
\label{simon}
\alpha = \left\{ \begin{array}{llll}
 6\times10^{-2}\left(r/{\rm AU}\right)^{-0.87} & \quad \ob > 0 & \& & r < r_{\rm outer} \\
 7.7\times10^{-4}\left(r/{\rm AU}\right)^{0.43} & \quad \ob < 0 & \& & r <  r_{\rm outer} \\
 3.3\times10^{-3} & \quad & & r \ge r_{\rm outer}
\end{array} \right.
\end{equation}

\noindent
where $r_{\rm outer} = 30$~AU.  The physical reason for assuming a single value for $\alpha$ (i.e., independent of magnetic field orientation) at  $r > r_{\rm outer}$ is the relatively weaker Hall effect at these large radii, thus removing the bimodal $\alpha$ behavior \citep{bai15,simon15b}.  We cap $\alpha$ at 0.05; larger values do not change our results significantly but require more radial zones for integrating the solution. Finally, we set $\alpha_{\rm mid} = 10^{-4}$ everywhere \cite[taken from recent simulations:][]{simon13a,bai15}, and therefore $\alpha$ is not allowed to go below $\alpha_{\rm mid}$.  

Solids will be concentrated towards the mid-plane and thus feel a diffusivity set by $\alpha_{\rm mid}$.  However,
since these solids are swept inwards by accretion, this ``viscosity" component is set by $\alpha$ as this parameterizes
the accretion flow.  This amounts to a definition of the ratio between diffusivity and accretion viscosity of $\xi = \alpha_{\rm mid}/\alpha$.

\subsection{Conditions for Planetesimal Formation}
\label{zcrit}

We consider planetesimal formation to be possible if one of two criteria are satisfied, First, following the arguments in \cite{youdin02}, if the metallicity increases beyond a critical value, it is possible for gravitational collapse of particles to proceed, independent of $\tau$ and without any interference from the Kelvin-Helmholtz instability.  We use Equation (15) in \cite{youdin02} to calculate a critical surface density for solids for each magnetic field orientation.  This can be easily converted to a critical metallicity, $Z_{\rm crit}$ and we assume that if $Z > Z_{\rm crit}$ prompt planetesimal formation occurs. 

If $Z < Z_{\rm crit}$, planetesimal formation is still possible when the values of $Z$ and $\tau$ fall within the allowed region of parameter space for the streaming instability \citep{youdin05} to operate.  This allowed space is still uncertain, but recently \cite{carrera15} carried out a large suite of streaming instability calculations and mapped out the allowed streaming region in $Z$--$\tau$ space.  In what follows, we include both planetesimal formation regions on plots of $Z$-$\tau$ space and assume that if the solution enters these regions, in situ planetesimal and planet formation occurs.

\section{Results}
\label{results}

Here, we run our model as described above and vary both the relative radial drift rate of solids, $\ratio$, and the fragmentation velocity, $v_{\rm frag}$, as these are both rather uncertain parameters.   We save a more in-depth parameter survey for future work.

\begin{figure}[ht!]
\centering
\includegraphics[width=0.5\textwidth]{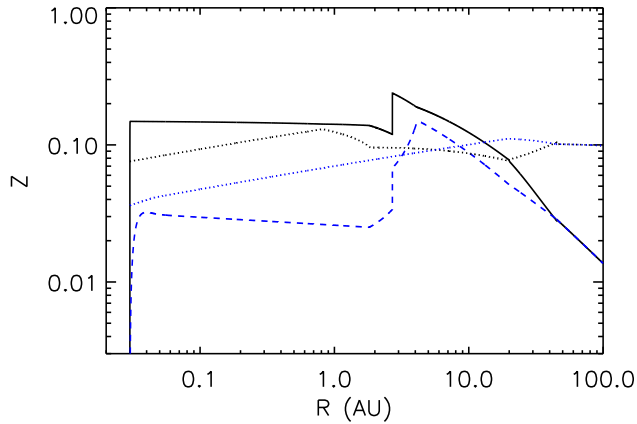}
\includegraphics[width=0.5\textwidth]{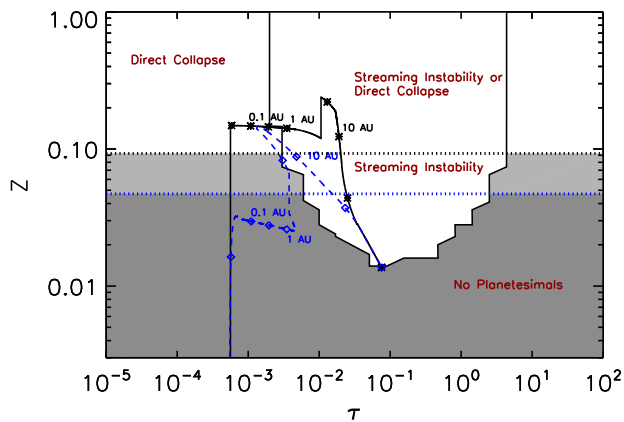}
\caption{Top: Steady state metallicity versus radius for the fiducial model with $\ob > 0$ (solid, black line) and $\ob < 0$ (dashed, blue line).  The
dotted lines are the critical metallicities above which direct collapse into planetesimals occurs \citep{youdin02}, with colors corresponding to the steady state model.  
Bottom:  The same information as in the top panel, but with solutions plotted in $Z-\tau$ space. The dotted lines are the critical metallicities averaged over $r < 0.3$~AU.  Radial information is marked on each curve with a spacing of 0.5 dex.  The grey regions correspond to no planetesimal formation (lighter grey applying only to $\ob > 0$), whereas the white regions allow planetesimal formation either through direct gravitational collapse or streaming instability followed by gravitational collapse (or both).  For $\ob > 0$, the metallicity is above the critical value at $r < 0.3$~{\rm AU}, whereas for $\ob < 0$, the metallicity does not exceed the critical value. 
}  
\label{fiducial}
\end{figure}

In Fig.~\ref{fiducial}, we show the fiducial case of $\ratio = 0.3$.  For $\ob < 0$, the metallicity at small radii does not enter either the region of direct gravitational collapse or streaming-initiated growth.   Instead, planetesimal appears to be possible at larger radii $r \gtrsim 3$~AU.  However, $\ob > 0$ shows regions $r \lesssim 0.3$AU have sufficiently high metallicity such that particles can directly collapse.  At larger radii, streaming-initiated growth can produce planetesimals. 

\begin{figure}[ht!]
\centering
\includegraphics[width=0.5\textwidth]{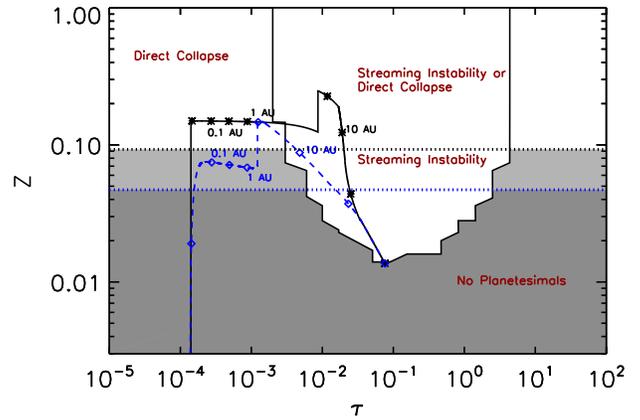}
\includegraphics[width=0.5\textwidth]{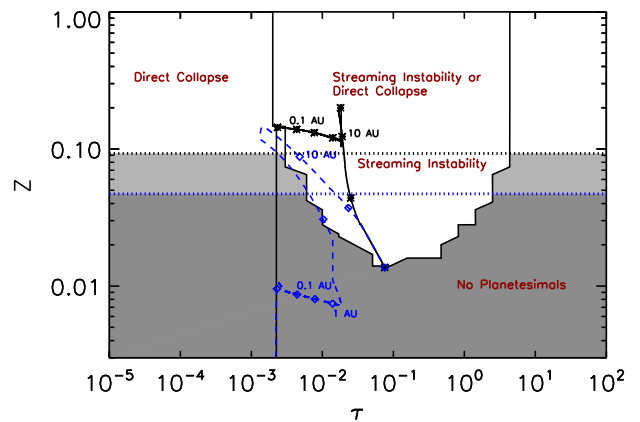}
\caption{Same as the bottom panel of Fig.~\ref{fiducial}, but with $v_{\rm frag} = 0.5$m/s (top) and $v_{\rm frag} = 2$m/s (bottom). In both cases, the $\ob > 0$ model allows for planetesimal creation, whereas this is only true
for $\ob < 0$ when $v_{\rm frag} = 0.5$m/s. Furthermore, there is always a large difference between the metallicities at small radii between the two models.}  
\label{vfrag_compare}
\end{figure}

\begin{figure}[ht!]
\centering
\includegraphics[width=0.5\textwidth]{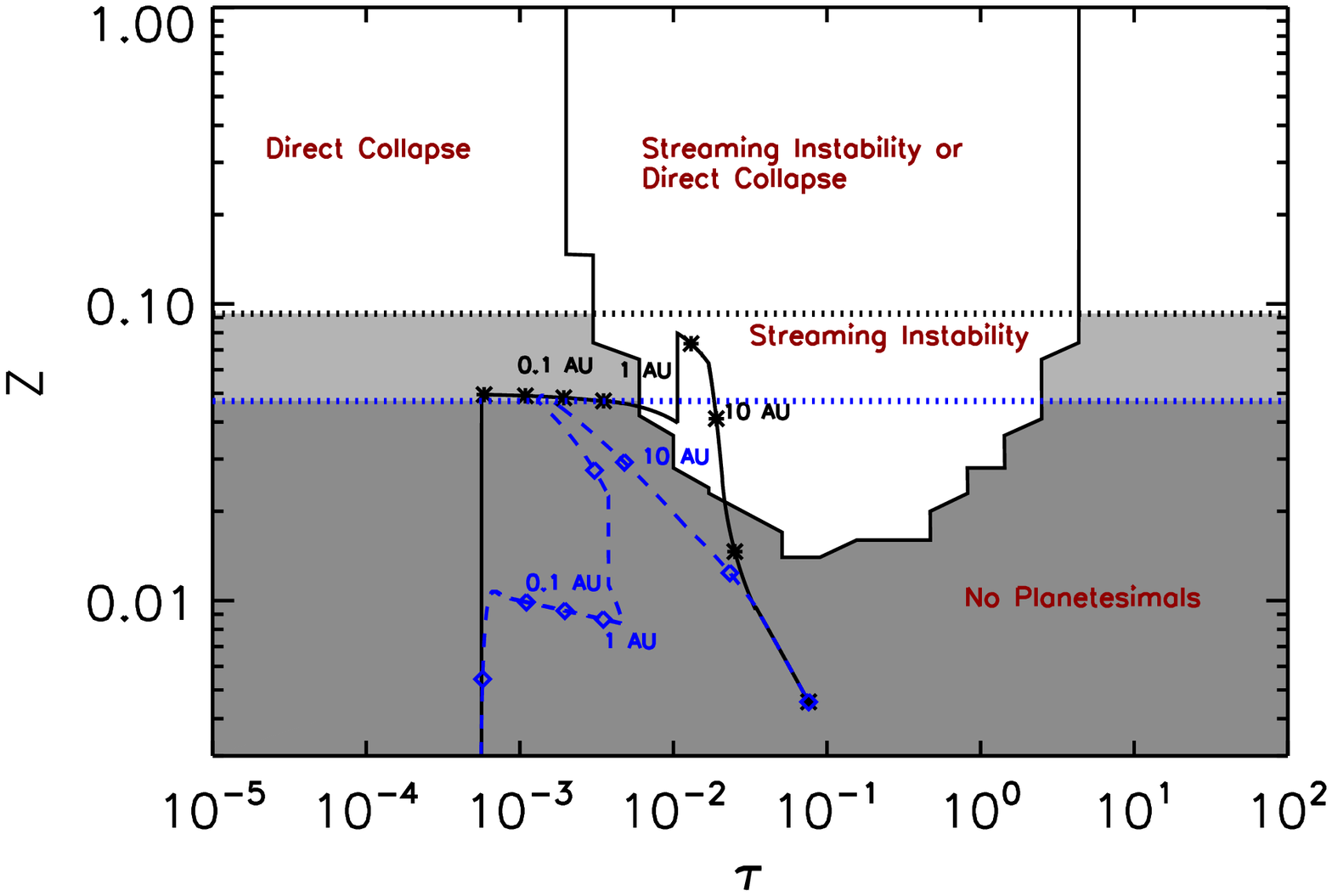}
\includegraphics[width=0.5\textwidth]{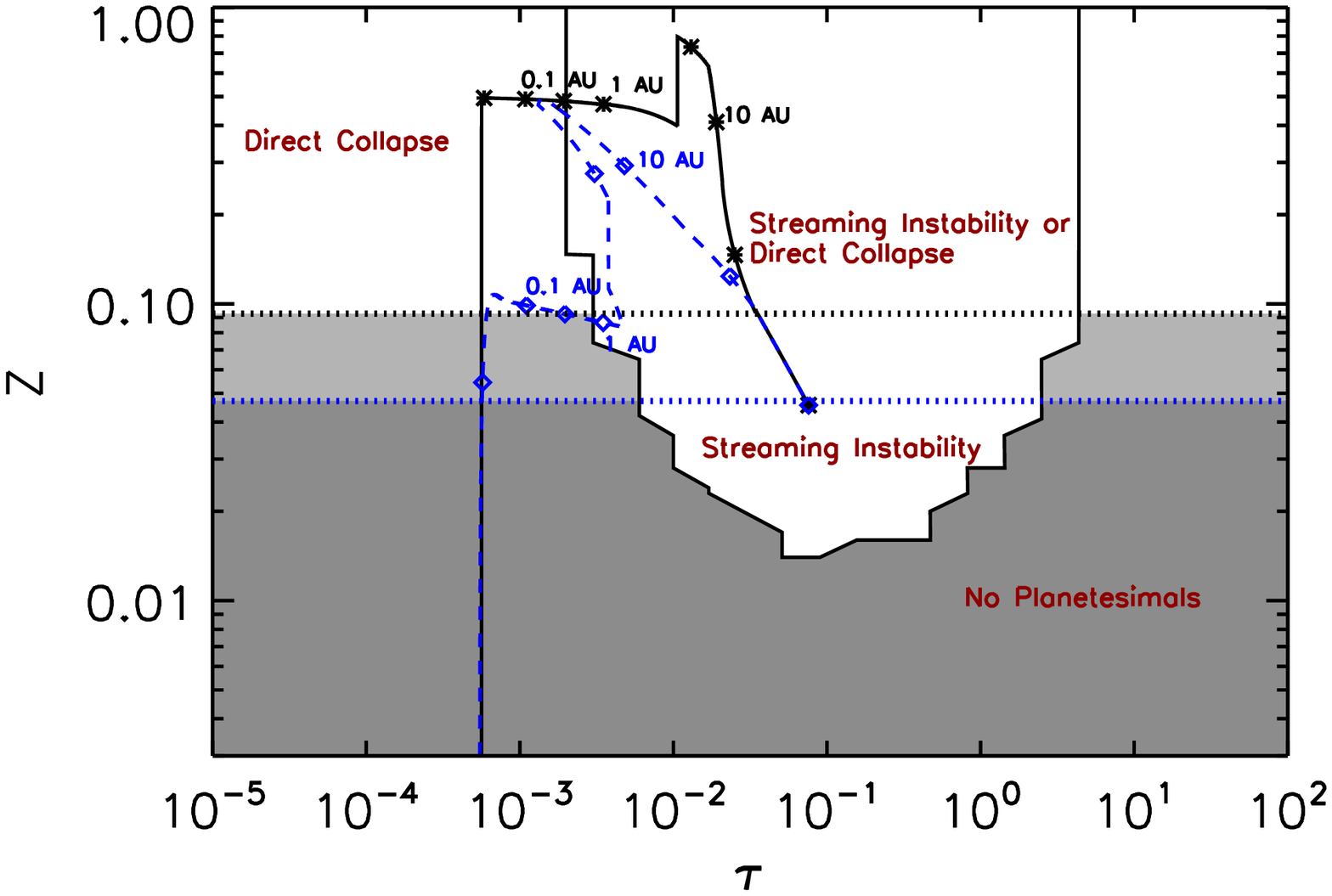}
\caption{Same as the bottom panel of Fig.~\ref{fiducial}, but with $\ratio = 0.1$ (top) and $\ratio = 1$ (bottom). The mass flux from large radial distances strongly influences both solutions.  However, there is always a large difference between the metallicities at small radii between the two field configurations. }  
\label{ratio_compare}
\end{figure}

We have also run our model with $v_{\rm frag} = 0.5$~m/s and $v_{\rm frag} = 2$~m/s; the output of these calculations is shown in Fig.~\ref{vfrag_compare}.  In all cases, the $\ob > 0$ solution allows for planetesimal formation. For $\ob < 0$ and  $v_{\rm frag} = 0.5$~m/s, the metallicity is larger than the critical value for collapse, whereas for the same field geometry and $v_{\rm frag} = 2$~m/s, the solution does not enter the region of planetesimal formation.   

The second parameter we explore is $\ratio$, and as explained above, we run our model for $\ratio = 0.1$ and 1 (shown in Fig.~\ref{ratio_compare}) in addition to the fiducial case (Fig.~\ref{fiducial}).  For the lowest flux case, neither magnetic field orientation enters the direct collapse regime, though the $\ob > 0$ case becomes unstable to the streaming instability between approximately 1 and 10 AU. The higher values of $\ratio$ again lead to $\ob > 0$ allowing for planetesimals through either direct collapse or streaming instability.  In the model with $\ratio = 1$, the $\ob < 0$ case is unstable to direct collapse at small radii, but enters the streaming unstable regime beyond $\sim 1$ AU.  We don't expect 
$\ratio$ to be significantly less than 0.1 as a value of $\ratio = 0.01$ would equate to co-accretion of solids with the gas flow at large distances from the star; a simple calculation reveals that radial drift strongly dominates accretion at these radii.  

Finally, we tested the robustness of the fiducial case to uncertainties in the gas surface density profile by using a different set of $\alpha$ values \cite[i.e., those from][]{bai15}. We found no qualitative difference. 

While the solutions fall into different regions of the parameter space, depending on the exact values for the parameters, the metallicity always reaches higher values for the $\ob > 0$ configuration compared to the $\ob < 0$ configuration.

\section{Discussion}
\label{summary}

\begin{figure}[ht!]
\centering
\includegraphics[width=0.5\textwidth]{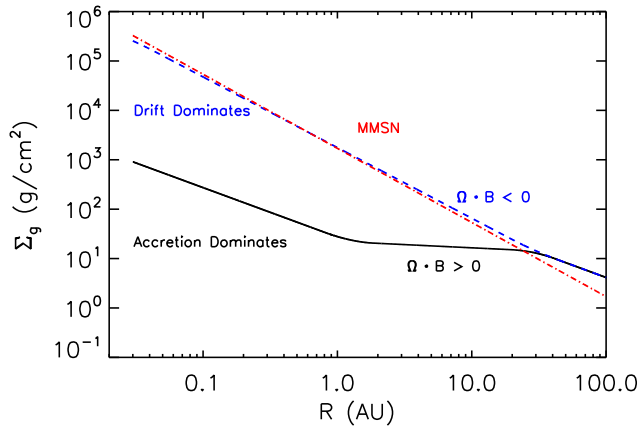}
\caption{Gas surface density as a function of radius for $\ob > 0$ (black, solid) and $\ob < 0$ (blue, dashed).  Also shown is the MMSN profile (red, dot-dashed).  As denoted on the plot, co-accretion
of solids with the gas flow dominates the inner regions of the disk for $\ob > 0$, whereas drift dominates in the $\ob < 0$ case. \vspace{0.1in}}  
\label{sigma}
\end{figure}

We have shown that whether or not a protoplanetary disk forms planets within the region $r < 0.3$ AU depends on the orientation of any large scale vertical magnetic field threading the disk with respect to the disk's rotation axis. 
The key ingredient is the dependence of the accretion stress on the product $\ob$, as mediated by the Hall effect. When $\ob > 0$, the resulting $\alpha$ values are enhanced and in a steady state, this
equates to a lower gas surface density. The opposite is true for $\ob < 0$; weaker $\alpha$ values lead to higher gas surface densities. 

It is worth pointing out that in the $\ob < 0$ case, the gas surface density depends more steeply on radius (see Fig.~\ref{sigma}) and in fact is reasonably close to the MMSN profile (which is particularly intriguing since
our own Solar System lacks tightly packed, close-in planets, in agreement with the $\ob < 0$ case).  Despite a larger radial drift velocity induced by this steeper profile, the shallower gas density profile still attains
higher solid concentration.  The reasons for this are two-fold.  First, with higher $\alpha$ values, the co-accretion of the solids with the gas dominates over radial drift, ensuring that (given a sufficiently high $\ratio$)
there will always be a relatively large number of solids at small radii, regardless of the value of $\tau$ (see Equation~(\ref{vp})). Second, the gas surface density is {\it lower} for $\ob > 0$, which means it is easier to attain relatively large $Z$.

In most of the cases explored here, $\ob > 0$ leads to sufficiently high metallicity for $r < 0.3$~AU to cause direct gravitational collapse into planetesimals, though the degree to which this is true depends on the value of $\ratio$.  Furthermore, the inner regions of the disk with $\ob < 0$ fall within the region of parameter space where planetesimal formation is difficult, though this result is even less robust as it depends on both $\ratio$ and $v_{\rm frag}$ (and possibly other parameters not explored here).   What is robust, however, is that for $\ob > 0$, the metallicity at small radii is enhanced, sometimes quite significantly, compared to $\ob < 0$.\footnote{Uncertainties in the various parameters may further obfuscate the difference between the solutions; more parameter exploration is required.}  Higher metallicity leads to more planetesimals 
and of larger sizes (Simon et al., in prep). Thus, even if parameters are such that both magnetic field orientations can lead to planetesimals at small radii, there remains a bimodal distribution in the number and sizes of these planetesimals.  Encouragingly, pebble accretion depends on the size of the accreting planetesimals (Kretke, private communication), and while one would need planet accretion and dynamical models to strengthen the link between
our results and exoplanet observations, the magnetic field geometry may very well be the key to explaining the dichotomy in exoplanet orbital configurations.

\acknowledgements

We thank Phil Armitage, Katherine Kretke, and Til Birnstiel for useful discussions,
and the referee, whose suggestions improved the quality of this work.
 We acknowledge support provided in part under contract with
the California Institute of Technology (Caltech) and the Jet Propulsion Laboratory (JPL) funded by NASA through the Sagan Fellowship Program executed by the
NASA Exoplanet Science Institute.

\end{document}